\documentclass[review]{AIAA}

\usepackage{float}                  
\usepackage{amsmath}           
\usepackage{amssymb, bm}   
\usepackage{graphicx}
\usepackage{booktabs}
\usepackage{setspace}

\begin{document}

\begin{spacing}{1.0}

\title{Waypoint Following Dynamics of a Quaternion Error Feedback Attitude Control System}

 \author{Mark~Karpenko\footnote{Research Associate Professor and corresponding author. e-mail: \tt mkarpenk@nps.edu }}
	\affiliation{\vskip 12pt Mechanical and Aerospace Engineering\\Naval Postgraduate School\\Monterey, CA, 93943, USA}

 \author{Julie~K.~Halverson\footnote{Lead Systems Engineer}}
	\affiliation{\vskip 12pt Space Sciences Mission Operations\\Goddard Space Flight Center\\Greenbelt, MD 20771, USA}

 \author{Rebecca Besser\footnote{Systems Engineer}}
	\affiliation{\vskip 12pt KBRWyle Government Services\\Huntsville,  AL, 35806, USA}

\begin{abstract}
Closed-loop attitude steering can be used to implement a non-standard attitude maneuver by using a conventional attitude control system to track a non-standard attitude profile. The idea has been employed to perform zero-propellant maneuvers on the International Space Station and minimum time maneuvers on NASA's TRACE space telescope. A challenge for operational implementation of the idea is the finite capacity of a space vehicle's command storage buffer. One approach to mitigate the problem is to downsample-and-hold the attitude commands as a set of waypoints for the attitude control system to follow. In this paper, we explore the waypoint following dynamics of a quaternion error feedback control law for downsample-and-hold. It is shown that downsample-and-hold induces a ripple between downsamples that causes the satellite angular rate to significantly overshoot the desired limit. Analysis in the $z$-domain is carried out in order to understand the phenomenon. An interpolating Chebyshev-type filter is proposed that allows attitude commands to be encoded in terms of a set of filter coefficients. Using the interpolating filter, commands can be issued at the ACS rate but with significantly reduced memory requirements. The attitude control system of NASA's Lunar Reconnaissance Orbiter is used as an example to illustrate the behavior of a practical attitude control system.

\end{abstract}

\maketitle

\end{spacing}

\section{Introduction}

A satellite attitude control system (ACS) is usually designed to implement a specific type of maneuver. For example, a quaternion error feedback control implements an Euler-axis slew in order to regulate to a desired quaternion target~\cite{Wie_book}. Closed-loop attitude steering is a concept for implementing maneuvers that are not compatible with a given ACS. Alternative maneuvers may be implemented by issuing a succession of attitude commands closely spaced in time to prompt the ACS to track the non-standard attitude trajectory. Closed-loop attitude steering has been used to implement the zero-propellant maneuver on the International Space Station~\cite{Bedrossian_2009} and to implement fast attitude maneuvers on NASA's TRACE space telescope~\cite{Karpenko_2014}. Other applications of closed-loop attitude steering include bright object avoidance~\cite{Lippman_2017}, reducing reaction wheel power consumption during slew~\cite{Marsh_2017} and recovering a spacecraft after a hardware failure~\cite{Kepler_2015, CMG_patent}. Closed-loop attitude steering is one option currently under consideration for implementing fast maneuvers on NASA's Lunar Reconnaissance Orbiter~\cite{LRO_breck}.

One issue that must be considered for operational implementation of closed-loop attitude steering is the finite capacity of the satellite command storage buffer, which may limit the number of commands that can be uplinked and subsequently issued to perform a particular slew. The memory requirement can be reduced by downsampling and holding the desired attitude trajectory with respect to the ACS sample time. Downsample-and-hold results in information loss by sampling the desired attitude trajectory at a lower rate. Downsample-and-hold is similar to representing an attitude trajectory as a series of waypoints that should be traversed by the ACS. In waypoint following, the attitude dynamics remain continuous, so in closed-loop attitude steering the satellite ACS behaves as a discrete-continuous control system.~It is logical then to question how the spacecraft will respond to such a sequence of inputs.

In this paper, the relationship between the downsample-and-hold frequency and the dynamics of a quaternion error feedback control system is explored in the context of closed-loop attitude steering. This is done by developing a discrete-continuous control system model and analyzing the waypoint following dynamics in the $z$-domain. It is shown that the resulting discrete time response does not adequately represent the dynamics between samples. In particular, downsample-and-hold is observed to induce a ripple between samples that can cause the satellite angular rate to significantly overshoot the ACS limit. This undesirable result can lead to violations of ACS monitors and cause the satellite to enter into a safe mode. In order to determine the magnitude of the inter-sample ripple, the discrete time model is augmented by using a modified $z$-transform~\cite{Franklin_1998} to enable the continuous time dynamics to be recreated in the discrete domain. Expressions relating the ripple magnitude to the commanding frequency and the control system gains are developed and it is seen that the ripple can only be reduced by increasing the frequency of downsample-and-hold. Doing so, however, can quickly exhaust the capacity of the command storage buffer. As an alternative to downsample-and-hold, an interpolating Chebyshev-type filter is proposed that allows attitude commands to be issued at the ACS rate but with significantly reduced memory requirements. The reduction in memory footprint is achieved by encoding the desired attitude profiles in terms of a set of filter coefficients. Simulation results for a fast attitude maneuver for the Lunar Reconnaissance Orbiter (LRO) are presented using a high-fidelity model of the LRO developed at the Goddard Space Flight Center. 

\section{Closed-Loop Attitude Steering}

This section illustrates the concept of attitude steering using NASA's Lunar Reconnaissance Orbiter (LRO) as an example of a practical attitude control system. The dynamics of the reaction wheel spacecraft and its attitude control system are summarized first. Then, it is shown how a fast attitude maneuver can be implemented on the the LRO by using a non-standard attitude trajectory to steer the closed-loop attitude control system.

\subsection{Attitude Control System Model}

The LRO is a three-axis stabilized spacecraft that uses a redundant array of four reaction wheels for attitude control. An appropriate model to describe the rotational dynamics of the satellite is given as
\begin{equation}
	\mathbf{ \dot x} = \left [ \begin{array}{c}
		\mathbf{Q}(\boldsymbol{\omega})\mathbf{q} \\
		\mathbf{J}^{-1}\left [- \boldsymbol{ \omega} \times (\mathbf{J}\boldsymbol{ \omega}  + J_{\text{w}}\mathbf{Z}\boldsymbol{\Omega}_{\text{w}} ) - \mathbf{Z}\mathbf{u} +\boldsymbol{\tau}_\text{ext} \right] \\
		J_{\text{w}}^{-1}\mathbf{u}
	\end{array} \right ]
	\label{eq_D06}
\end{equation}
where the state vector is taken as $\mathbf{x}= \left [\mathbf{q}|\boldsymbol{\omega}|\boldsymbol{\Omega}_{\text{w}}\right ]^T$. The $4\times1$ vector $\mathbf{q}$ gives the attitude quaternions ($q_4$ is taken as the scalar), $3\times1$ vector $\boldsymbol{\omega}$ gives the spacecraft angular rates and, and $4\times1$ vector $\boldsymbol{\Omega}_{\text{w}}$ gives the reaction wheel speeds. The $4\times1$ control vector is $\mathbf{u}=\boldsymbol{\tau}_{\text{w}}$, where $\boldsymbol{\tau}_{\text{w}}$ are the reaction wheel control torques, $J_{\text{w}}$ is the inertia of the wheel rotors, $\mathbf{Z}$ is a $3\times4$ column matrix of unit vectors relating the individual wheel spin axes to the body-fixed frame, and $\mathbf{h}=J_{\text{w}}\mathbf{Z}\boldsymbol{\Omega}_{\text{w}}$ is the reaction wheel angular momentum in the body-fixed frame. The external torque, due to gravity gradient and solar-radiation pressure, is denoted as $\boldsymbol{\tau}_\text{ext}$. The skew-symmetric matrix, $\mathbf{Q}(\boldsymbol{\omega})$, is given as

\begin{equation}
	\mathbf{Q}(\boldsymbol{\omega}) = \dfrac{1}{2}\left [ 		
	\begin{array}{cccc}
		0 & \omega_3 & -\omega_2 & \omega_1 \\
		-\omega_3 & 0 & \omega_1 & \omega_2 \\
		\omega_2 & -\omega_1 & 0 & \omega_3 \\
		-\omega_1 & -\omega_2 & -\omega_3 & 0 \\				
	\end{array}
	\right ]
	\label{eq_D02}
\end{equation}

LRO's observing mode attitude control law, like many spacecraft, is based on quaternion error feedback~\cite{Calhoun_2007}. The body control torque, $\boldsymbol{\tau}_{\text{c}}$, is given as
\begin{equation}
\boldsymbol{\tau}_{\text{c}} = \mathbf{J}\left(K_p\mathcal{N}(2\mathbf{a}_e)-K_r\boldsymbol{\omega}+K_i\int2\mathbf{a}_e\right) + \boldsymbol{\omega}\times\left(\mathbf{J}\boldsymbol{\omega}+J_{\text{w}}\mathbf{Z}\boldsymbol{\Omega}_{\text{w}}\right)
\label{ACS_01}
\end{equation}
where $K_p=0.057$, $K_i=0.0023$, and $K_r=0.4$ are the underdamped proportional-integral-rate control gains (see~[\cite{Calhoun_2007}]). Variable $\mathbf{a}_e$ is the vector part of the error quaternion, i.e. $\mathbf{q}_e=[\mathbf{a}_e, s_e]^T$. Only the error quaternion and its integral are used to drive the system. Feedback of the spacecraft angular rate vector provides damping, because the desired angular rate vector for a step input is the null vector. During a slew, the integral gain is zeroed to avoid overshoot resulting from integral feedback and activated when the attitude error drops below a certain threshold. In~\eqref{ACS_01}, the error quaternion computed as
\begin{equation}
\mathbf{q}_e = \mathbf{q}^{-1}\otimes\mathbf{q}_{\text{tgt}}
\label{ACS_02}
\end{equation}

The control law includes a gyroscopic compensation term as well as a proportional limiter, $\mathcal{N}(\mathbf{a}_e)$, on the quaternion error that enforces a slew rate limit while simultaneously preserving the attitude vector direction. The proportional limiter is not needed for closed-loop attitude steering since any slew rate constraints are accounted for as part of the design of the maneuver. The command torques for the individual reaction wheels are determined by a Moore-Penrose pseudoinverse allocation scheme
\begin{equation}
\boldsymbol{\tau}_{\text{w}} = \mathbf{Z}^T\left(\mathbf{Z}\mathbf{Z}^T\right)^{-1}\boldsymbol{\tau}_{\text{cmd}} =\mathbf{Z}^{\#}\boldsymbol{\tau}_{\text{cmd}} 
\label{ACS_03}
\end{equation}

Normally, the quaternion target, $\mathbf{q}_\text{tgt}$, in~\eqref{ACS_02} is fixed and the satellite ACS simply regulates to the desired attitude by zeroing the attitude error. In order to implement a non-standard maneuver, however, the quaternion target is a function of time, i.e. $\mathbf{q}_\text{tgt}(t)$, that is ideally sampled at the ACS update rate.

\subsection{Closed-Loop Attitude Steering: Fast Attitude Maneuver}

A simple example is now presented to illustrate the overall concept of closed-loop attitude steering. In this example, an attitude command trajectory has been generated to perform a fast attitude maneuver for the LRO. The fast attitude maneuver is solved as a dynamic optimization problem that maximizes the instantaneous momentum-to-inertia ratio during the slew. The details of the fast maneuver design, which can reduce the overall slew time by more than 45\% as compared to a standard slew plan, are presented elsewhere~(see ~[\cite{LRO_breck}]).  

\begin{figure}[!htbp]
\begin{center}
\begin{tabular}{cc}
	\includegraphics[width=3.0in,clip]{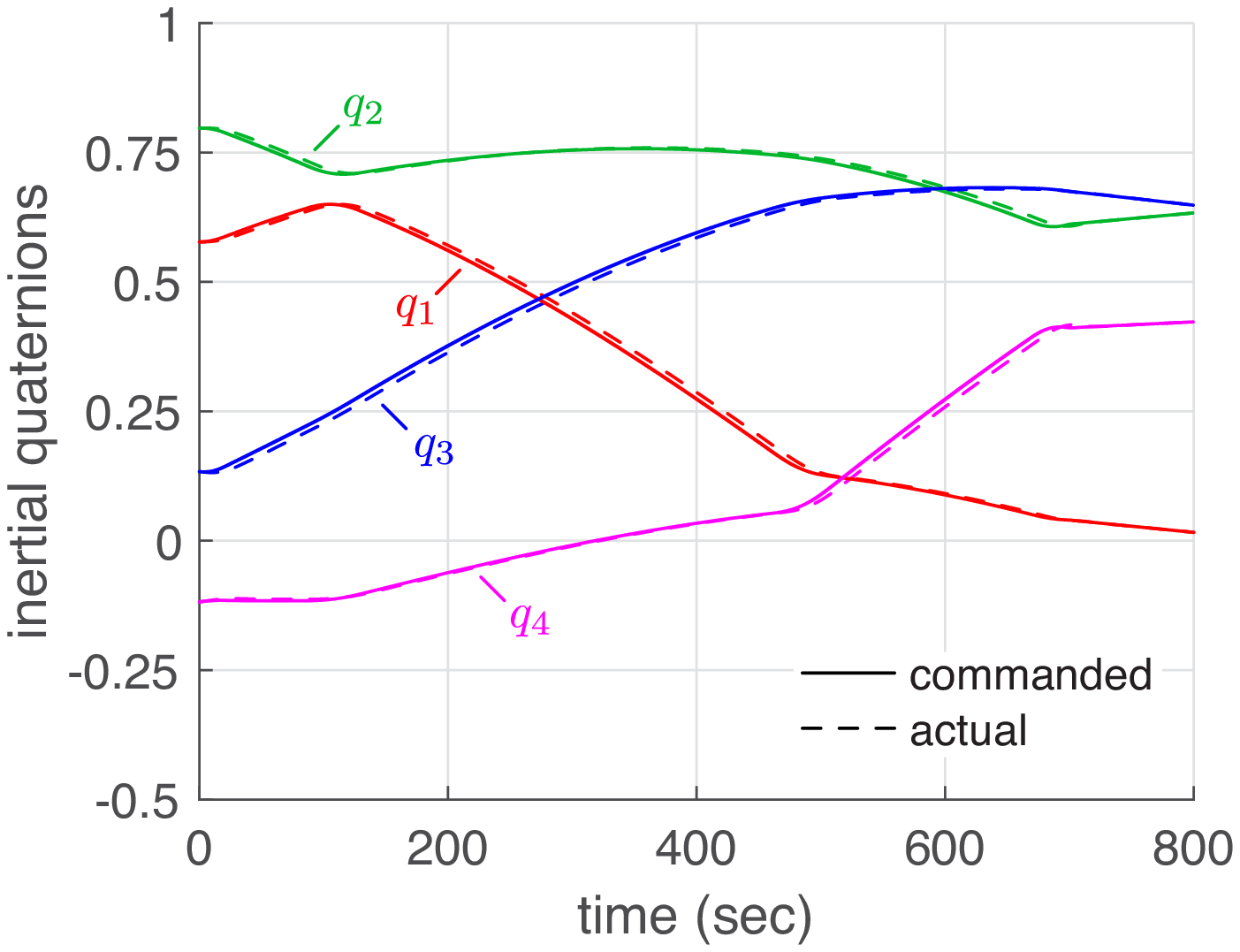}  & 	\includegraphics[width=3.0in,clip]{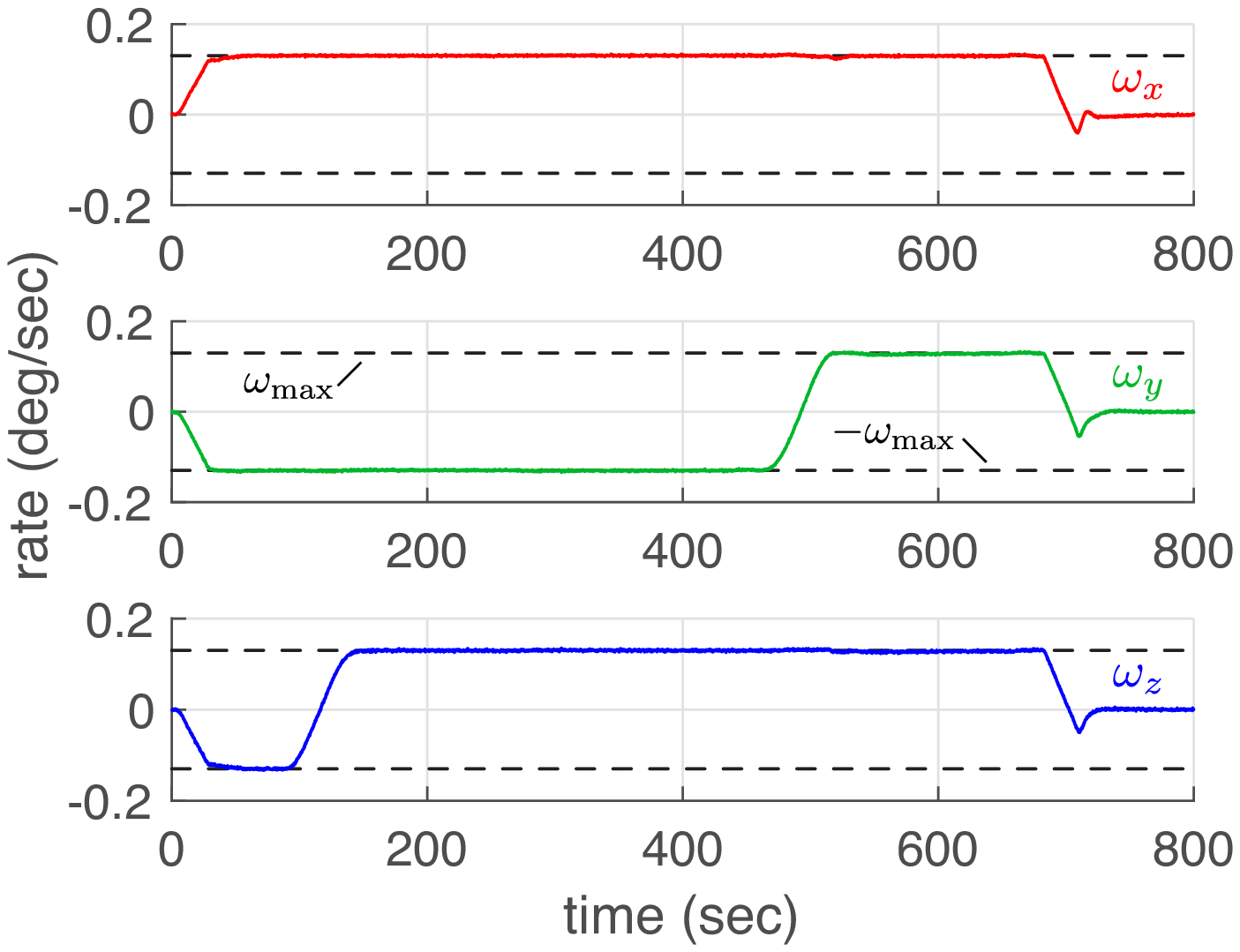} \\
\small{(a)} & \small{(b)} \\
	\includegraphics[width=3.0in,clip]{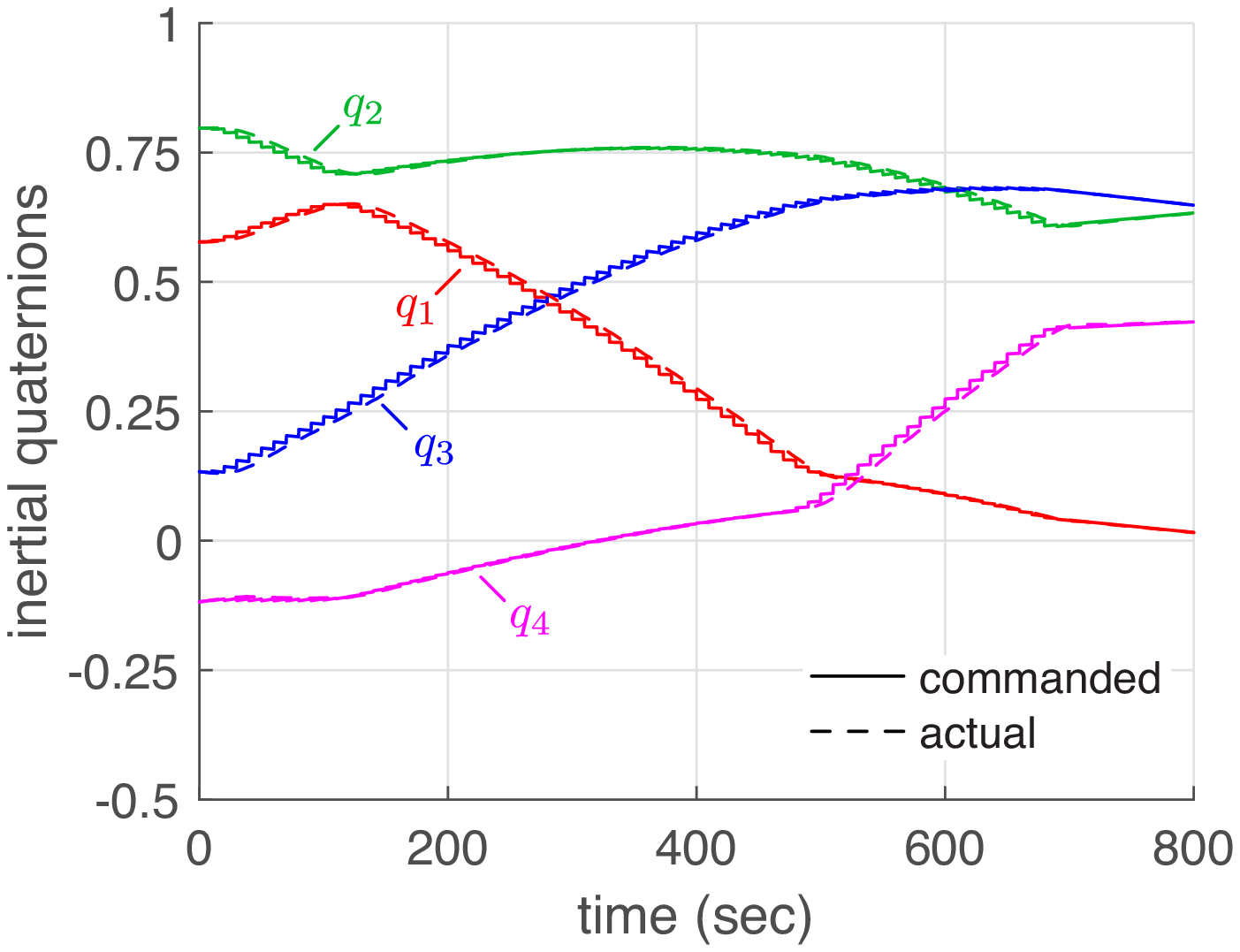}  & 	\includegraphics[width=3.0in,clip]{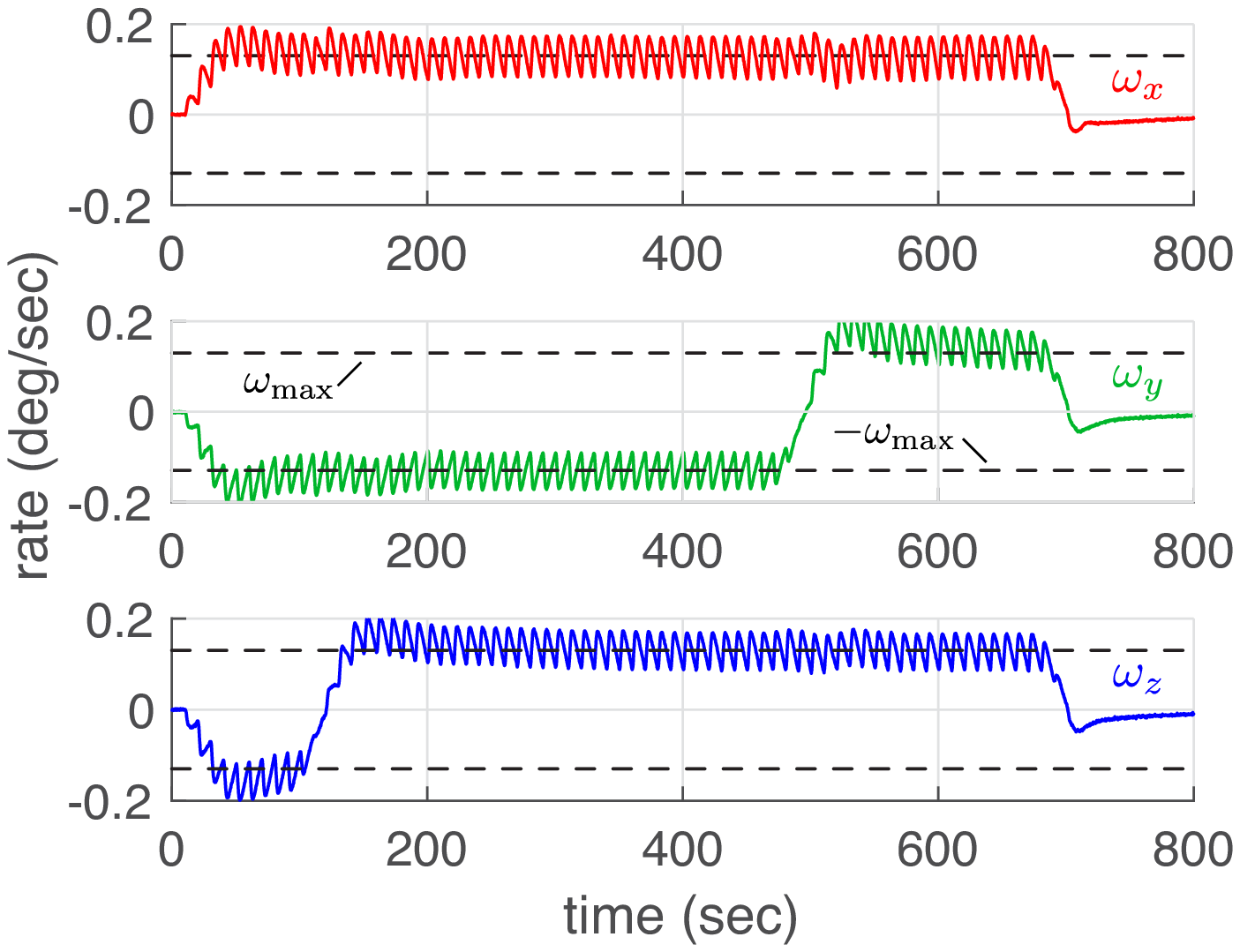} \\
\small{(c)} & \small{(d)} \\
\end{tabular}
\caption{Simulation of a fast attitude maneuver designed for the Lunar Reconnaissance Orbiter: (a) attitude quaternions for waypoint sampling at the ACS frequency (5~Hz); (b) body rates for waypoint sampling at the ACS frequency (5~Hz); (c) attitude quaternions for downsampled waypoints (0.1~Hz); (d) body rates for downsampled waypoints (0.1~Hz).}
\label{Fig_01}
\end{center}
\end{figure}

The fast maneuver was implemented in a high-fidelity simulation of the LRO's attitude control system by sample-and-hold of the desired attitude trajectory for various frequencies. The results are shown in Figure~\ref{Fig_01}. Figures~\ref{Fig_01}a and~\ref{Fig_01}b shows the simulation results for sample-and-hold of the attitude command trajectory at the ACS rate ($T=0.2$ sec). As can be seen, the spacecraft attitude quaternions slightly lag the commanded profiles. This is to be expected since the integral gain of the quaternion error feedback control law is set to zero during slew. The angular rate profiles (given in the body-fixed frame) show a rotation with rate buildup to, but never exceeding, LRO's limiting value of 0.13 deg/sec along all three axes. The timed transitions between the negative and positive rate limits observed for the $y$ and $z$-axes are a result of optimizing the momentum-to-inertia ratio during the slew to the target attitude. 

Due to the slew rate constraint of the LRO, the maneuver time for the fast attitude maneuver is still quite large (708 sec as compared to 1112 sec for the standard slew). Thus, implementing closed-loop attitude steering at the ACS rate requires more than 3500 time-tagged commands to be uplinked and stored in the spacecraft command buffer. Assuming that the data is stored as floats, the memory footprint required for storing the attitude command samples is about 70-kilobytes (100-bytes per slew second). A single fast attitude maneuver can therefore consume nearly 50\% of the available command storage capacity available on a typical spacecraft.

The memory footprint can be reduced by implementing downsample-and-hold to reduce the number of waypoints used to represent the  the fast attitude maneuver trajectory. For example, performing downsample-and-hold at $T=10$ sec, the number of command that need to be uplinked and stored can be reduced to about 70. The associated memory footprint is now only 1.4-kilobytes, which is manageable. The simulation results for downsample-and-hold ($T=10$ sec) is shown in Figures~\ref{Fig_01}c and~\ref{Fig_01}d. A cursory inspection of the attitude quaternions, Figure~\ref{Fig_01}c, seems to indicate that the maneuver can be performed successfully when the attitude commands are sampled at a reduced rate. However, Figure~\ref{Fig_01}d shows that downsample-and-hold introduces a large ripple that significantly degrades the angular rate response. The ripple is a result of the dynamics introduced by the zero-order-hold on the otherwise continuous time system. The rate variations are undesirable because the variations in the angular rate are also manifest as variations in the individual reaction wheel speeds (not shown), which increases the duty-cycle of the wheels. Moreover, the large peak value of the angular rates can lead to violations of ACS flight software monitors and potentially cause the satellite to enter into a safe mode. In the following, we seek to characterize this undesirable ripple in terms of the ACS parameters and the downsampling time in order to better understand the origins of the phenomenon.

\section{Waypoint Following Dynamics for Downsample-and-Hold}

\begin{figure}[!htbp]
\begin{center}
\includegraphics[width=5.5in,clip]{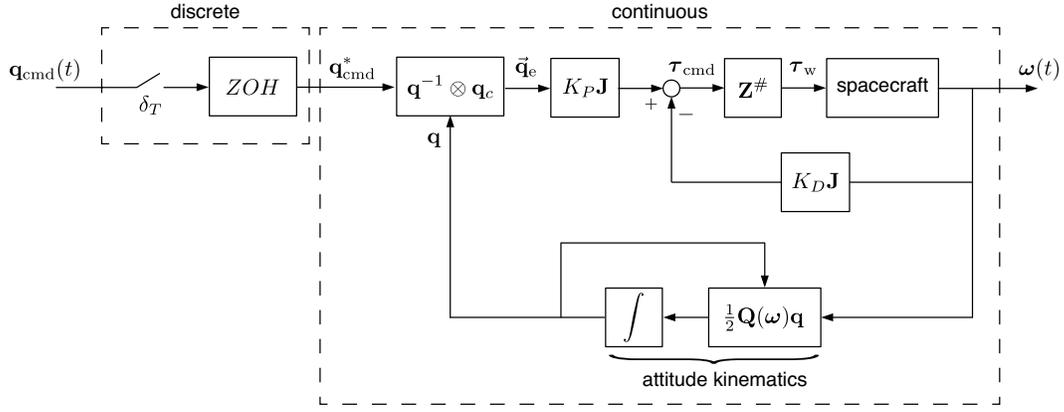}
\caption{Block diagram of discrete-continuous time attitude control system.}
\label{Fig_02}
\end{center}
\end{figure}

In this section, the waypoint following dynamics are analyzed for the case where attitude command trajectories are downsampled with respect to the ACS update frequency. As such, the behavior of the attitude control system can be thought of in terms of continuous-time processing of a discrete-time input signal. The resulting discrete-continuous system may be modeled and analyzed as an equivalent discrete-time control system. A block diagram of the discrete-continuous attitude control system is shown in Figure~\ref{Fig_02}.

\begin{figure}[!htbp]
\begin{center}
\begin{tabular}{c}
\includegraphics[width=4.75in,clip]{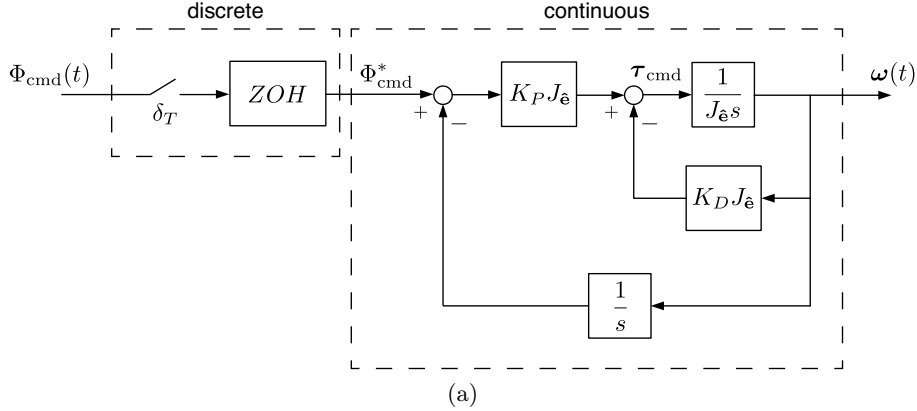}\\
\small{(a)} \\
\\
\includegraphics[width=4.35in,clip]{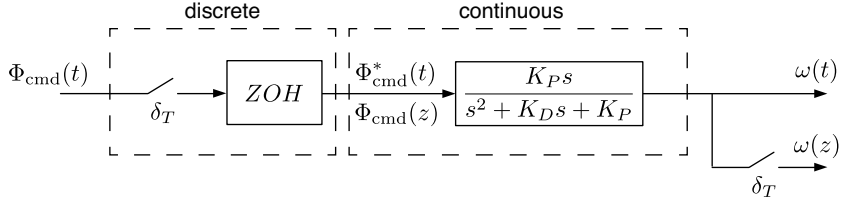}\\
\small{(b)} \\
\end{tabular}
\caption{Block diagram of discrete-continuous time attitude control system: (a) as a second-order system; (b) reduced for $z$-transformation. }
\label{Fig_03}
\end{center}
\end{figure}

Quaternion error feedback control logic with feed forward of the gyroscopic coupling terms effectively decouples the roll, pitch and yaw dynamics~\cite{Calhoun_2007}. Thus, it is sufficient to analyze the characteristics of an individual control channel. The block diagram of Figure ~\ref{Fig_02} can therefore be simplified down to a second-order system (see Figure~\ref{Fig_03}). In Figure~\ref{Fig_03}, $J_{\mathbf{\hat e}}$, is the inertia about the rotational axis (roll, pitch, or yaw) and $\Phi_{\text{cmd}}$ is the rotation angle about the same axis. The discrete-time transfer function of the block diagram for the reduced discrete-continuous time attitude control system in Figure~\ref{Fig_03}b is
\begin{equation}
\dfrac{\omega(z)}{\Phi_\text{cmd}(z)}=\dfrac{z-1}{z}\mathcal{Z}\left\{\dfrac{K_Ps}{s(s^2+K_Ds+K_P)} \right\}
\label{eq_w00}
\end{equation}
where the term $(z-1)/z\cdot\mathcal{Z}\{1/s\}$ arises due to the zero-order hold on the input.

To evaluate \eqref{eq_w00}, begin with the factored form of the continuous plant $P(s)/s=ab/(s+a)(s+b)$, which has the inverse Laplace transform (for zero initial conditions)
\begin{equation}
p(t) = \dfrac{ab}{b-a}\left(e^{-at}-e^{-bt}\right)
\label{eq_w02}
\end{equation}
Re-writing~\eqref{eq_w02} at instants $t=kT$ gives
\begin{equation}
p(kT) = \dfrac{ab}{b-a}\left(e^{-akT}-e^{-bkT}\right)
\label{eq_w03}
\end{equation}
Taking the $z$-transform of~\eqref{eq_w03} we have
\begin{equation}
\mathcal{Z}\left\{p(kT)\right\} = \dfrac{ab}{b-a}\left(\dfrac{1}{1-z^{-1}e^{-aT}}-\dfrac{1}{1-z^{-1}e^{-bT}}\right)
\label{eq_w04}
\end{equation}
Finally, the transfer function of the reduced system may be written as 
\begin{equation}
\dfrac{\omega(z)}{\Phi_\text{cmd}(z)}=\dfrac{ab}{b-a}\cdot\dfrac{z-1}{z}\cdot\dfrac{(e^{-aT}e^{-bT})z}{(z-e^{-aT})(z-e^{-bT})}
\label{eq_w05}
\end{equation}
The specific values of the gains of the LRO attitude control system have been designed for an underdamped second-order response with $\omega_n =\sqrt{K_P}= 0.24$ rad/sec and $\zeta=K_D/(2\sqrt{K_P})=0.85$. For the underdamped case, the continuous time poles are located at, $a=-\zeta\omega_n+j\omega_d$ and $b=-\zeta\omega_n-j\omega_d$, where $\omega_d=\omega_n\sqrt{1-\zeta^2}$ is the damped natural frequency and $j=\sqrt{-1}$. Substituting the complex conjugate poles in~\eqref{eq_w05} and simplifying the result gives
\begin{equation}
\dfrac{\omega(z)}{\Phi_\text{cmd}(z)}=\dfrac{\tfrac{\omega_n}{\sqrt{1-\zeta^2}}e^{-\zeta\omega_nT}\sin(\omega_dT)(z-1)}
{z^2-2e^{-\zeta\omega_nT}\cos(\omega_dT)z+e^{-2\zeta\omega_nT}}
\label{eq_w06}
\end{equation}

Since the desired response of the attitude control system is a rate-limited slew, angle $\Phi_{\text{cmd}}(t)$ is essentially a ramp signal with magnitude equivalent to the slew rate limit, $\omega_\text{max}=0.13$ deg/sec. In the discrete-time domain the ramp input is given as 
\begin{equation}
\Phi_\text{cmd}(z)=\omega_\text{max}T\dfrac{z}{(z-1)^2}
\label{eq_w07}
\end{equation}
Since, in general~\eqref{eq_w05} is not unit gain, the ramp input should be adjusted by a factor $K$ to normalize the response and ensure that $\lim_{k\to\infty}\omega[k]=\omega_{\max}$ as required. With this in mind, the rate response in the discrete-time domain may be written as
\begin{equation}
\omega(z)=\omega_\text{max}TK\dfrac{\tfrac{\omega_n}{\sqrt{1-\zeta^2}}e^{-\zeta\omega_nT}\sin(\omega_dT)z}
{(z-1)(z^2-2e^{-\zeta\omega_nT}\cos(\omega_dT)z+e^{-2\zeta\omega_nT})}
\label{eq_w08}
\end{equation}

\begin{figure}[!htbp]
\begin{center}
\begin{tabular}{cc}
\includegraphics[width=3.0in,clip]{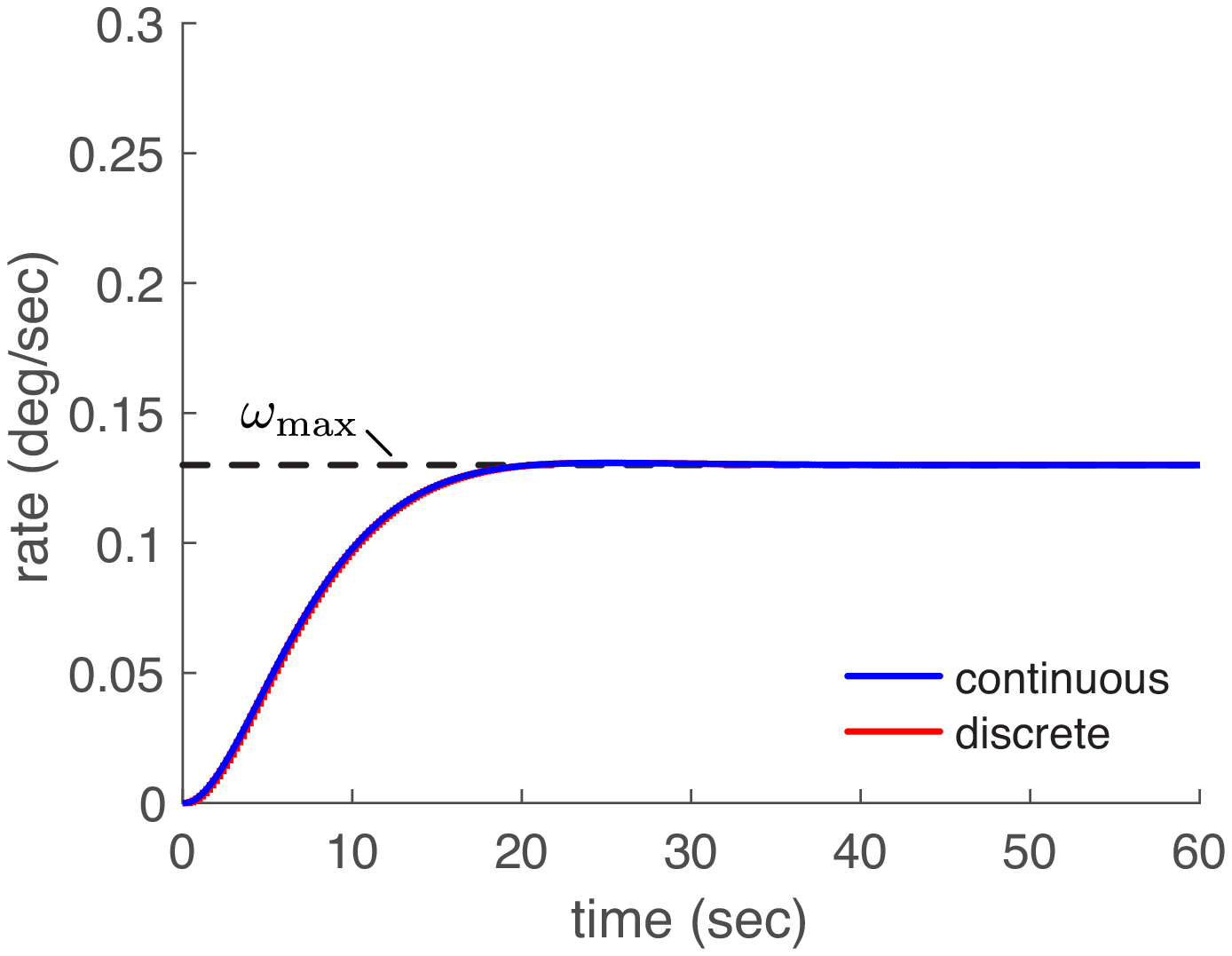} & \includegraphics[width=3.0in,clip]{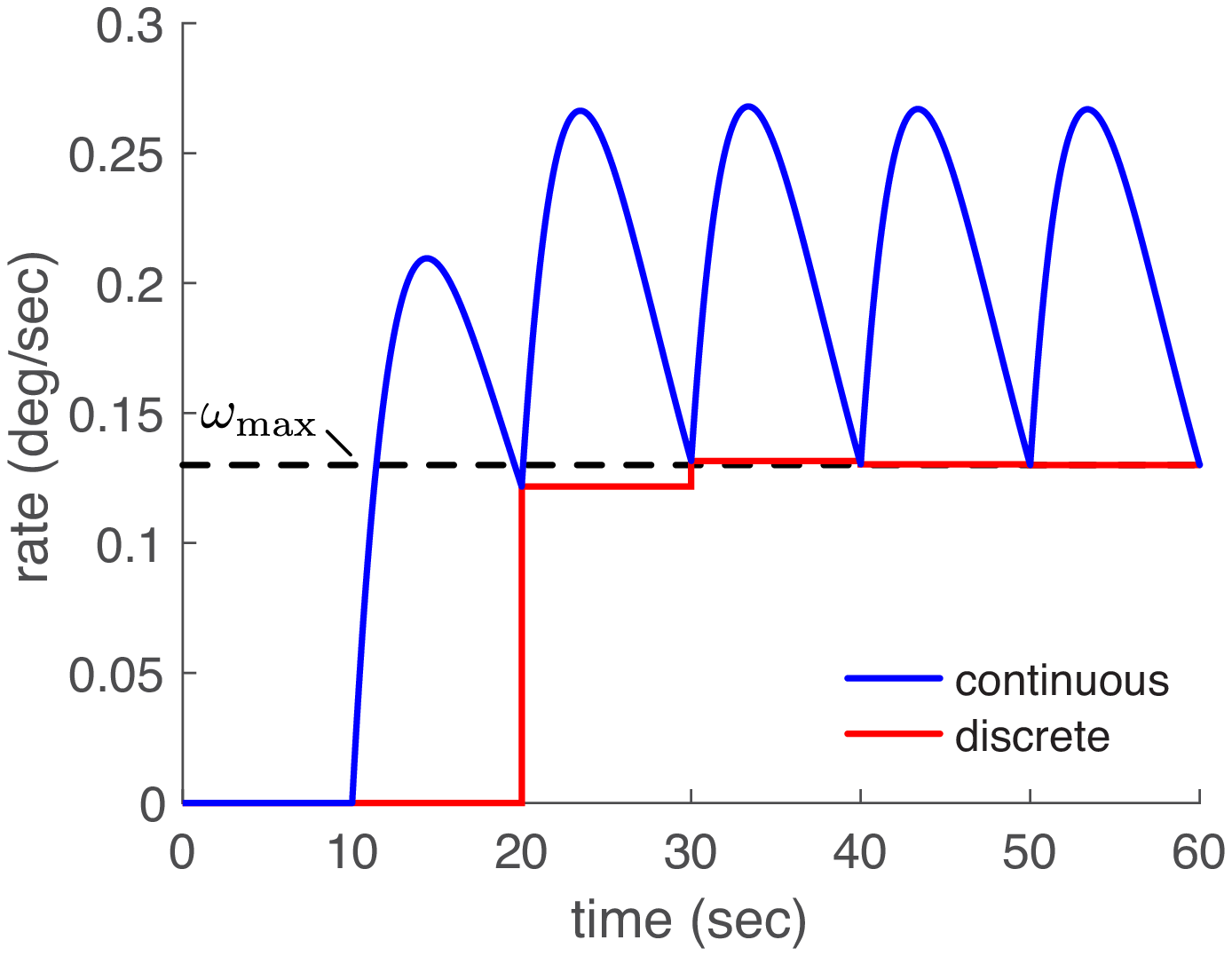}\\
\small{(a)} & \small{(b)}
\end{tabular}
\caption{Slew rate response from simulation of the discrete time attitude control system~\eqref{eq_w08}: (a) sample-and-hold at $T=0.2$ sec and (b) downsample-and-hold at $T=10$~sec.}
\label{Fig_04}
\end{center}
\end{figure}

Simulations of \eqref{eq_w08} give the slew rate responses shown in Figure~\ref{Fig_04} for sample-and-hold at the ACS rate (5~Hz) and at the downsample-and-hold rate (0.1~Hz).  The continuous-time responses of the single-channel attitude control system (which are consistent with the spacecraft simulation of Figure~\ref{Fig_01}) are superimposed for reference. Referring to Figure~\ref{Fig_04}, it is clear that the description of the waypoint following dynamics provided by~\eqref{eq_w08} properly models the response of the continuous control system for sample-and-hold at the ACS rate (Fig~\ref{Fig_04}a). However for downsample-and-hold, Fig~\ref{Fig_04}b shows that the discrete time model~\eqref{eq_w08}, does not predict the ripple behavior of the continuous attitude control system because~\eqref{eq_w08} assumes the output is held between samples.

\subsection{Intersample Ripple in the $z$-Domain}

\begin{figure}[!htbp]
\begin{center}
\includegraphics[width=5in,clip]{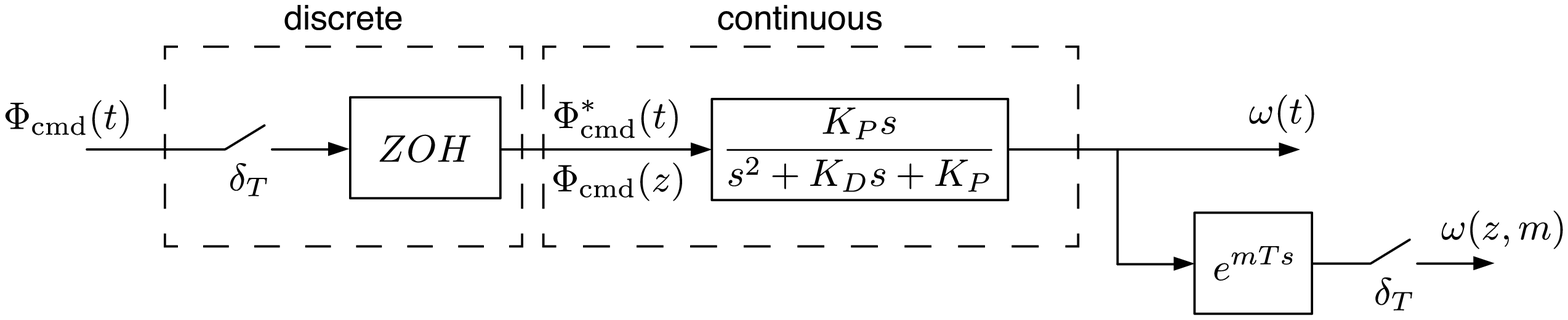}\\
\caption{Block diagram of time-shifted discrete-continuous time attitude control system. }
\label{Fig_time_shifted_TF}
\end{center}
\end{figure}

In order to model the system output between samples of the input, we may insert a bank of time-shifted samplers at the output of the system~\cite{Franklin_1998}. The effect of time-shifted sampling can be evaluated by inserting a transport delay, $e^{-\lambda s}$, where $\lambda = lT-mT$, at the output of the plant as shown in Figure~\ref{Fig_time_shifted_TF}. The associated modified $z$-transform may be written as \begin{equation}
\dfrac{\omega(z,m)}{\Phi_\text{cmd}(z)}=\dfrac{z-1}{z}\cdot\dfrac{1}{z^l}\cdot\mathcal{Z}\left\{\dfrac{P(s)}{s}e^{mTs} \right\}\qquad 0\leq m <1
\label{eq_z00}
\end{equation}
where $l$-delays are introduced to ensure that the transfer function is causal. Upon the evaluation of~\eqref{eq_z00}, the output of the plant will then be represented by samples taken at instants $t=(k+m)T$ instead of $t=kT$ as are given by~\eqref{eq_w06}.

\begin{figure}[!htbp]
\begin{center}
\includegraphics[width=3.5in,clip]{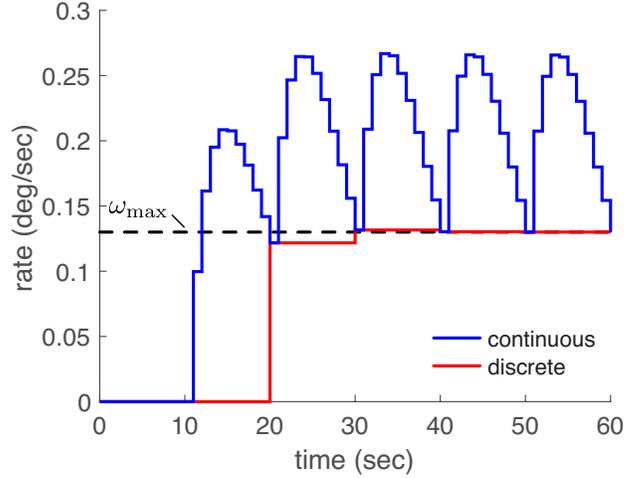}
\caption{Reconstruction of continuous-time slew rate response using time-shifted samples for downsample-and-hold at $T=10 sec$.}
\label{Fig_05}
\end{center}
\end{figure}

Re-writing~\eqref{eq_w03} at instants $t=(k+m)T$ gives
\begin{equation}
p(kT+mT) = \dfrac{ab}{b-a}\left(e^{-a(kT+mT)}-e^{-b(kT+mT)}\right)
\label{eq_z01}
\end{equation}
The $z$-transform of~\eqref{eq_z01} is
\begin{equation}
\mathcal{Z}\left\{p(kT+mT)\right\} = \dfrac{ab}{b-a}\left(\dfrac{e^{-amT}}{1-z^{-1}e^{-aT}}-\dfrac{e^{-bmT}}{1-z^{-1}e^{-bT}}\right)
\label{eq_z02}
\end{equation}
Using~\eqref{eq_z02}, equation \eqref{eq_z00} can be written as
\begin{equation}
\dfrac{\omega(z,m)}{\Phi_\text{cmd}(z)}=\dfrac{ab}{b-a}\cdot\dfrac{z-1}{z^{1+l}}\cdot\dfrac{(e^{-amT}-e^{-bmT})z^2+(e^{-aT}e^{-bmT}-e^{-bT}e^{-amT})z}{(z-e^{-aT})(z-e^{-bT})}\qquad 0\leq m <1
\label{eq_z03}
\end{equation}
with $l=1$ (i.e. the first time-shifted sample does not occur until $t=(1+m)T$). Since the LRO attitude control system is underdamped,~\eqref{eq_z03} may be re-written as
\begin{equation}
\dfrac{\omega(z,m)}{\Phi_\text{cmd}(z)}=\dfrac{\tfrac{\omega_n}{\sqrt{1-\zeta^2}}\left(e^{-\zeta\omega_nTm}\sin(\omega_dTm)z+e^{-\zeta\omega_nT(1+m)}\sin(\omega_dT(1-m))\right)(z-1)}
{z^2-2e^{-\zeta\omega_nT}\cos(\omega_d)z+e^{-2\zeta\omega_nT}} \qquad 0\leq m <1
\label{eq_z04}
\end{equation}
Now, by simulating the bank of time-shifted samplers, i.e. by varying the value of $m\in [0,1)$, it is possible to recreate the continuous time rate ripple as shown in Figure~\ref{Fig_05}, but using a $z$-domain transfer function. Thus, the nature of~\eqref{eq_z04} may be studied in order to better understand the origin of the large variations in the rate output that is observed for a downsample-and-hold implementation.

\begin{figure}[!htbp]
\begin{center}
\includegraphics[width=3.5in,clip]{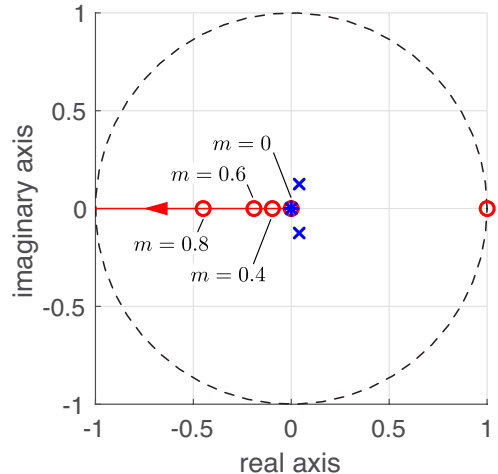}
\caption{Pole-zero pattern of  $\omega(z,m)/\Phi_\text{cmd}(z)$ showing locus of additional zero for various values of $m$ for downsample-and-hold at $T=10$ sec.}
\label{Fig_06}
\end{center}
\end{figure}

Comparing, the time-shifted $z$-transform of~\eqref{eq_z04} with the original $z$-transform~\eqref{eq_w06}, it is evident that the inserted transport delay introduces a zero whose location is a function of $m$, in addition to the original zero at $z=1$ that is associated with the zero-order hold. The locations of two poles, however, remain unchanged. Moreover, the transfer function gain is altered due to the appearance of parameter $m$ in the numerator polynomial. The pole-zero pattern of the modified $z$-transform for $T=10$ sec is illustrated in Figure~\ref{Fig_06} for various values of time-delay parameter $m$. As is seen, the locus of the zero location associated with the time-delay moves from the origin and to the left and along the negative real axis as the value of $m$ increases. To determine the effect of the additional zero, it is also necessary to consider the equivalent $s$-domain damping and natural frequency of the discrete time poles. The equivalent natural frequency of a discrete time pole is given by 
\begin{equation}
\bar \omega_n = \left \vert \dfrac{\ln(z)}{T} \right \vert 
\label{eq_z05}
\end{equation}
and the equivalent damping is 
\begin{equation}
\bar \zeta = -\cos(\angle\ln(z))
\label{eq_z06}
\end{equation}
For the pole-zero pattern of Figure~\ref{Fig_06}, we have $\bar \omega_n=0.24$ rad/sec and $\bar \zeta =0.85$, i.e. the continuous-time natural frequency and  damping ratio of the satellite attitude control system.  For these values, the number of samples per oscillation is $N=2\pi/(\bar \omega_nT\sqrt{1-\bar \zeta^2})\approx 5$ so the transient will die out in 5 samples as shown in Figure~\ref{Fig_06}. Moreover, for $N=5$ and $\bar \zeta = 0.707$, is it has been shown by~Franklin \emph{et al.}~\cite{Franklin_1998} that the presence of the additional zero (associated with the time-shifted sampling) will have very little influence on the peak value of the rate response since the additional zeros lie on the negative real axis. Hence, we may conclude that the appearance of the additional zero is not the reason for the large ripple observed in the attitude control system rate response. 

\begin{figure}[!htbp]
\begin{center}
\includegraphics[width=3.5in,clip]{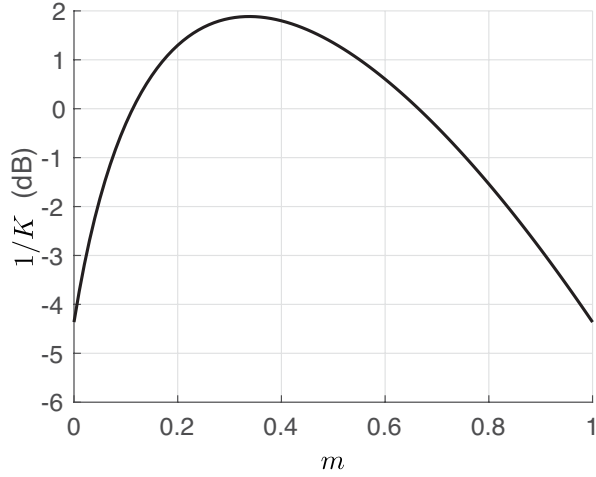}
\caption{Gain factor of $\omega(z,m)/\Phi_\text{cmd}(z)$ for various values of $m$  for downsample-and-hold at $T=10$ sec.}
\label{Fig_07}
\end{center}
\end{figure}

The variation in the normalizing gain factor $K$ (assuming a ramp input) is plotted in decibels for various values of $m$ in~Figure~\ref{Fig_07}. The plot reveals that as the value of $m$ increases from 0 to 1, the value of the normalizing gain varies by about 6-dB. This implies that the DC-gains of the time-shifted transfer functions vary by a factor of approximately 2. This variation correlates well to the magnitude of the rate ripple observed in Figures~\ref{Fig_04}b and equivalently~Figure~\ref{Fig_05}. We may thus conclude that the variation in the DC-gain of the $z$-domain transfer function is the primary cause for the undesirable continuous-time ripple response. 

In order to relate the ripple magnitude to the to parameters, $\omega_n$ and $\zeta$, of the attitude control system and to the attitude command sampling interval, $T$, the discrete-time final value theorem may be utilized. For a unit ramp-input, the steady-state rate response is given by
\begin{equation}
\begin{split}
\lim_{k\to\infty} \omega(k,m) = \lim_{z\to1}&(z-1)\omega(z,m) \\&=\dfrac{KT\tfrac{\omega_n}{\sqrt{1-\zeta^2}}\left(e^{-\zeta\omega_nTm}\sin(\omega_dTm)+e^{-\zeta\omega_nT(1+m)}\sin(\omega_dT(1-m))\right)}{1-2e^{-\zeta\omega_nT}\cos(\omega_d)+e^{-2\zeta\omega_nT}}
\label{eq_z07}
\end{split}
\end{equation}
To determine the shifted time associated with the ripple peak, it is necessary to find the fractional sample time, $m$, that maximizes~\eqref{eq_z07}. Setting the partial derivative of \eqref{eq_z07} with respect to $m$ equal to zero gives a candidate for the maximizing value, $m^\ast$, as
\begin{equation}
m^\ast = \dfrac{\ln\left(b(1-e^{-aT})\right)+\ln\left(a(e^{-bT}-1)\right)}{(a+b)T}
\label{eq_z08}
\end{equation}
 
Table~\ref{Table_02} lists the peak steady-state rate ripple response for the spacecraft attitude control system as a function of the downsample-and-hold interval. The results were obtained by substituting the value of $m^\ast$ obtained from~\eqref{eq_z08} into~\eqref{eq_z07} and evaluating the final value. As is seen, the peak ripple magnitude decreases as the time between input samples is decreased. This is to be expected because decreasing the downsampling time reduces the information loss associated with downsample-and-hold. Since the violations of the ACS monitors may be triggered when the rate exceeds a predetermined threshold, the satellite may enter into a safe mode if the peak ripple is not maintained below an acceptable level. As indicated by the results of Table~\ref{Table_02}, reducing the peak ripple can only be done by decreasing the downsample-and-hold frequency of the attitude command inputs. This, however, increases the demands placed on the spacecraft command storage buffer. For example, to maintain the ripple peak at the desired slew rate limit of $0.13$ deg/sec, the input commands must be downsampled and held at least once per second. This translates to a command storage rate of 20-bytes per slew second. Thus, for the 708 second fast attitude maneuver of Figure~\ref{Fig_01}, about 14-kilobytes would still be needed to implement the slew. Clearly, an alternative approach is needed for managing the data storage requirements for closed-loop attitude steering if the concept is to be instantiated on an operational spacecraft such as the LRO.

\begin{table}[!htb]
\caption{Peak rate magnitude for various downsample-and-hold frequencies.}
\centering
\begin{tabular}{ccc}
\hline 
$T$ (sec) & $m^\ast$ & $\omega_{\text{peak}}$ (deg/sec) \\
\hline
10 & 0.34 & 0.268 \\
5 & 0.42 & 0.156 \\
2 & 0.47 & 0.134 \\
1 & 0.48 & 0.132 \\
0.2 & -- & 0.131 \\
\hline
\label{Table_02}
\end{tabular}
\end{table}

\section{Back to the Basis: A Chebyshev-Like Interpolating Filter}

As described earlier in this paper, fast attitude maneuvers are obtained by solving a dynamic optimization problem that maximizes the instantaneous momentum-to-inertia ratio during the slew. In references~[\cite{Karpenko_2014,LRO_breck}], Pseudospectral optimal control theory is utilized to solve the problem. See reference [\cite{IFAC_survey}] for an overview of the approach.~In a pseudospectral method, orthogonal polynomial interpolants of the form
\begin{equation}
x(\tau)=\sum_{j=0}^{N}\phi_j(\tau)x_j,\quad -1\leq\tau\leq1
\label{eq_PS00}
\end{equation}
are utilized to interpolate the state trajectories. In~\eqref{eq_PS00}, $\phi_j(\tau)$ are the $N$th-order Lagrange interpolating polynomials
\begin{equation}
\phi_j(\tau) = \prod_{i=0,i\neq j}^N\dfrac{\tau-\tau_i}{\tau_j-\tau_i}
\label{eq_PS01}
\end{equation}
where $\tau_i$ and $\tau_j$ are elements of a set of $N$ distinct nodes on the interval $\tau \in [-1,1]$.

For a Chebyshev pseudospectral method, the nodes are the Chebyshev-Gauss-Lobatto (CGL) nodes given by~[\cite{Fahroo_2002}]
\begin{equation}
\tau_j =-\cos(\pi j/N), \quad j=0,\hdots,N
\label{eq_PS04}
\end{equation}
Referring to~\eqref{eq_PS04}, the CGL nodes are not uniformly spaced in $[-1,1]$. Instead, they are clustered around the endpoints of the computational interval. This clustering avoids the Runge phenomenon normally associated with high-order polynomial interpolation on a uniform grid.~Lagrange polynomials and the CGL nodes may be utilized in order to develop an interpolation filter that can be used to reconstruct information about the fast maneuver trajectories that would otherwise be lost if the attitude commands were downsampled and processed through a conventional sample and hold logic. 

To illustrate the overall concept, consider the finite impulse response $h(\tau)$ of a conventional Chebyshev interpolating filter for a single polynomial segment over $\tau\in[-1,1]$. The impulse response is given by~\cite{Babic_2013}
\begin{equation}
h(\tau) = \sum_{j=0}^Nc_jT_j(\tau)
\label{filter_00}
\end{equation}
In~\eqref{filter_00}, which has a similar structure to~\eqref{eq_PS00}, the basis functions $T_j(\tau)$ are Chebyshev polynomials of the first kind instead of Lagrange polynomials. Thus, $T_0(\tau)=1$, $T_1(\tau)=\tau$, $T_2(\tau)=2\tau^2-1$, $T_3(\tau)=4\tau^3-3\tau$, and so on. Coefficients $c_j$ of~\eqref{filter_00} are the adjustable filter parameters. In digital signal processing, these filter coefficients are normally ``designed'' (see~[\cite{DSP_book}]) to achieve certain filter characteristics, e.g. passband, stopband, etc. In this paper, however, we utilize the concept of the filter coefficients as a means to encode information pertaining to the shapes of attitude command trajectories over a slew. This can be done because the shape of the waveform to be produced is known~\emph{a priori}. The slew trajectory can then be reproduced as the finite impulse response of the filter. To utilize the conventional Chebyshev interpolating filter for this purpose, the $N$th-order filter coefficients can be determined using the standard equation
\begin{equation}
c_j = \dfrac{2}{N}\sum_{k=0}^Nx(\tau_k)T_j(\tau_k)
\label{filter_01}
\end{equation}
where $x(\tau_k)$ is a sample of command trajectory, $x$, taken at the instants, $\tau_k$. 

In the conventional Chebyshev filter, instants $\tau_k$ are given as
\begin{equation}
\tau_k = \cos\left(\dfrac{\pi(k-1/2)}{N}\right), \quad k=1,\hdots,N
\label{filter_011}
\end{equation}
From~\eqref{filter_011}, it is apparent that sample times $\tau_k$ are not the same as the sample times of the CGL nodes given by~\eqref{eq_PS04}. Thus, the application of a conventional Chebyshev interpolating filter for reconstructing a fast maneuver trajectory requires post processing in order to re-sample the pseudospectral solution (given at the CGL nodes) to the Chebyshev filter's grid. Moreover in the conventional Chebyshev filter, the interpolating polynomials are Chebyshev polynomials and not Lagrange polynomials as employed by the pseudospectral method. As a consequence, the response of the conventional Chebyshev filter may not faithfully represent the desired attitude trajectories.

To sidestep these issues, the $N$th-order Chebyshev polynomial, specifically its time derivative, may be mapped to produce Lagrange polynomials and sample times consistent with a Chebyshev pseudospectral method as described in~[\cite{Fahroo_2002}]
\begin{equation}
\phi_j(\tau) = \dfrac{(-1)^{j+1}}{N^2a_j}\cdot\dfrac{(1-\tau^2)\dot T_N(\tau)}{\tau-\tau_j}
\label{filter_02}
\end{equation}
where
\begin{equation}
a_j = 
\begin{cases}
2,\qquad j=0,N \\
1,\qquad 1\leq j\leq N-1
\end{cases}
\end{equation}

\begin{figure}[!htbp]
\begin{center}
\includegraphics[width=3.5in,clip]{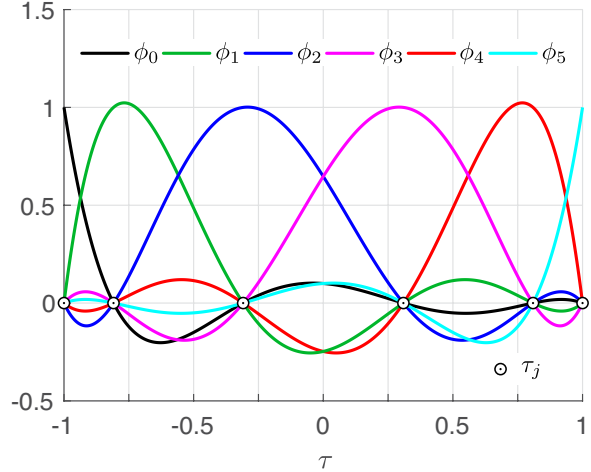}
\caption{Sixth-order Lagrange polynomials.}
\label{Fig_09}
\end{center}
\end{figure}

The finite impulse response of the Chebyshev-like interpolating filter~\eqref{filter_02} for a single polynomial segment over $[-1,1]$ may be defined as 
\begin{equation}
h(\tau)=\sum_{j=0}^{N}\bar c_j\dfrac{(-1)^{j+1}}{N^2a_j}\cdot\dfrac{(1-\tau^2)\dot T_N(\tau)}{\tau-\tau_j}
\label{filter_03}
\end{equation}
where the filter coefficients $\bar c_j$ (distinguished from $c_j$ of the conventional Chebyshev filter) are determined directly from the desired attitude state(s) at the CGL nodes, i.e. $\bar c_j = x(t_j)$. This is evident from Figure~\ref{Fig_09}, which shows that all except the $j$-th Lagrange polynomial are zero at the CGL points and that the value of the $j$-th polynomial is unity at its CGL point. 

We note that~\eqref{filter_03} may be alternatively implemented in terms of the Barycentric interpolation formula. The Barycentric interpolation formula (see~[\cite{Berrut_2004}]) is
\begin{equation}
h(\tau)=\dfrac{\sum_{j=0}^N\dfrac{w_j}{\tau-\tau_j}\bar c_j}{\sum_{j=0}^N\dfrac{w_j}{\tau-\tau_j}}
\label{filter_001}
\end{equation}
where $w_j$ are the Barycentric weights given by
\begin{equation}
w_j=\dfrac{1}{\prod_{i\neq j}(\tau_j-\tau_i)},\quad j=0,\hdots,N
\label{filter_002}
\end{equation}
An advantage of using~\eqref{filter_001} is that the Barycentric weights are constants that may be precomputed and stored to reduce the computational load associated with evaluating the interpolant.

The are several additional advantages of utilizing the modified Chebyshev interpolating filter described above in place of conventional sample and hold logic for closed-loop attitude steering. First, because~\eqref{filter_03} is an interpolation formula, it is possible to obtain an interpolated command input at any instant in time. Thus, inputs to the spacecraft ACS can be generated at the ACS update rate in order to circumvent issues with rate ripple induced by downsample-and-hold. Second, because the description of the desired attitude trajectories are encoded in terms of the filter coefficients, only the filter coefficients need to be uplinked to the spacecraft in order to generate the interpolated commands, provided that the Chebyshev interpolating filter with generic coefficients is instantiated as part of the ACS logic. As will be seen in the next section, this approach can significantly reduce the storage footprint associated with closed-loop attitude steering. Third, because the polynomial basis of the new Chebyshev interpolating filter is consistent with the polynomial basis utilized to design the non-standard attitude maneuver, command trajectories for closed-loop attitude steering may be reproduced without loss of information at any servo rate. This later aspect is crucial to ensure that any operational constraints considered during the maneuver design are satisfied when the maneuver is implemented via closed-loop attitude steering.

\section{Attitude Steering Using An Interpolation Filter}

In this section we illustrate how the modified Chebyshev filter described in the previous section may be utilized to improve the tracking dynamics of a quaternion error feedback attitude control system for closed-loop attitude steering.

\subsection{Ramp Response}

In the preceding analysis, a ramp command was employed to study the waypoint following dynamics of the quaternion error feedback attitude control system. The interplay between the continuous-time dynamics of the attitude control system and the command sampler introduced a ripple on the rate response for downsample-and-hold commanding. Since the Chebyshev-like filter developed in the previous section performs an interpolation operation between the command samples (now taken at the CGL points), it is expected that the response of the attitude control system will be much improved when the filter is employed.

In Figure~\ref{Fig_04}b, six evenly spaced downsamples ($T=10$ sec) of the ramp input were applied to the attitude control system over the time interval $t\in[0,60]$ sec. In order to apply the Chebyshev-like interpolation filter, the ramp input must be sampled at the non-uniform CGL points over $t\in[0, 60]$ so that the coefficients of the finite impulse response filter can be determined. The computation is straightforward for the Chebyshev-like filter. Six filter coefficients for the ramp input are listed in Table~\ref{Table_cj}. Because Lagrange interpolating polynomials are used, the values for the coefficients of the modified filter are the same as the values of the ramp signal at the CGL sampling instants.The interpolating filter can reproduce the ramp input precisely at any instant of time $t\in[0, 60]$ including at the ACS sampling instants. 

\begin{table}[!htb]
\caption{Filter coefficients for interpolating a ramp input with modified Chebyshev filter (N=6).}
\centering
\begin{tabular}{ccc}
\hline 
$t_j$ (sec)  & $\tau_j$ & $\bar c_j$ \\
\hline
0.0		& -1.0000 & 0.0 \\ 
5.73		& -0.8090 & 0.74\\
20.73	& -0.3090 & 2.69\\
39.27	& 0.3090 & 5.11\\
54.27	& 0.8090 &7.06\\
60.00	& 1.0000 & 7.80\\
\hline
\label{Table_cj}
\end{tabular}
\end{table}

\begin{figure}[!htbp]
\begin{center}
\begin{tabular}{cc}
\includegraphics[width=3.0in,clip]{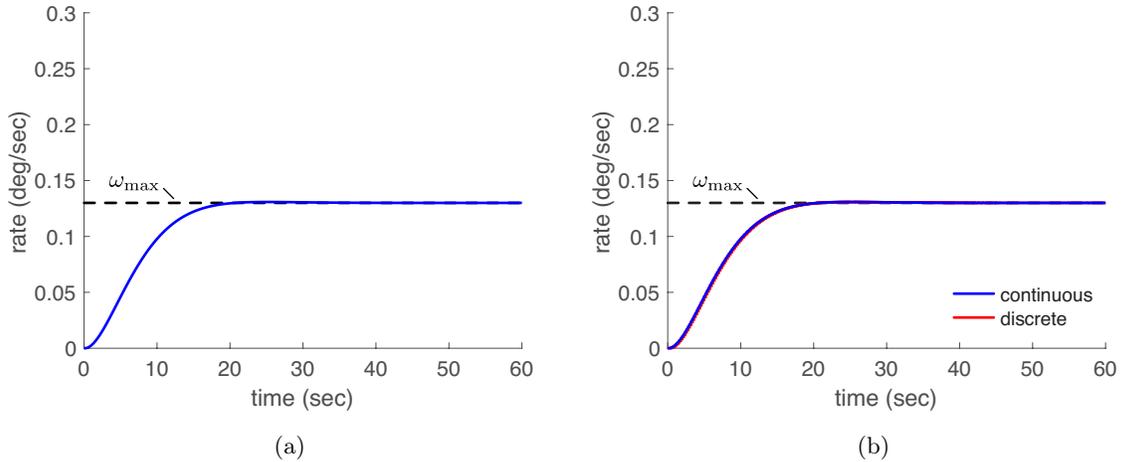} & \includegraphics[width=3.0in,clip]{T02_sampled_rate.eps}\\
\small{(a)} & \small{(b)}
\end{tabular}
\caption{Continuous-time slew rate response from simulation of the attitude control system using: (a) proposed interpolation filter and (b) sample-and-hold at the ACS rate (Figure~\ref{Fig_04}b repeated).}%
\label{Fig_10}
\end{center}
\end{figure}

A time simulation of the quaternion error feedback attitude control system was performed using the modified Chebyshev filter. The rate response is given in Figure~\ref{Fig_10}. The rate response for downsample-and-hold at $T=10$ sec, i.e. Figure~\ref{Fig_04}b, is also repeated for comparison. In each case, only six samples of the attitude input signal are used. In the case of the modified Chebyshev filter the samples are taken at the CGL points whereas for downsample-and-hold, the samples are taken uniformly at $T=10$ sec intervals. As can be seen, using the interpolating filter to generate the input to the ACS does not induce the inter-sample ripple associated with downsample-and-hold. In fact, the rate response is identical to what is obtained for sample-and-hold at the ACS rate.

\subsection{Fast Attitude Maneuver}

As a means to further illustrate the efficacy of the modified Chebyshev filter, a bank of four interpolating filters was implemented in the high fidelity simulation of the LRO spacecraft and the fast attitude maneuver of Figure~\ref{Fig_01} was simulated. Because the attitude profile for the fast maneuver is not as simple as a ramp input, additional filter coefficients are required in order properly encode the quaternion commands. For the fast attitude maneuver of Figure~\ref{Fig_01} it was determined that 50 filter coefficients are needed. As before, these were determined directly from the quaternion profiles by sampling at the CGL points. The coefficient data along with the slew time interval are all that needs to be uplinked and stored on the satellite in order to implement the fast maneuver using the filter bank. The associated memory footprint for implementing the fast attitude maneuver is reduced from 70-kilobytes for sample-and-hold at the ACS to 1-kilobyte using the interpolating filter.

\begin{figure}[!htbp]
\begin{center}
\begin{tabular}{cc}
	\includegraphics[width=3.0in,clip]{Hz02_HiFi_quat.eps}  & 	\includegraphics[width=3.0in,clip]{Hz02_HiFi_rate.eps} \\
\small{(a)} & \small{(b)} \\
	\includegraphics[width=3.0in,clip]{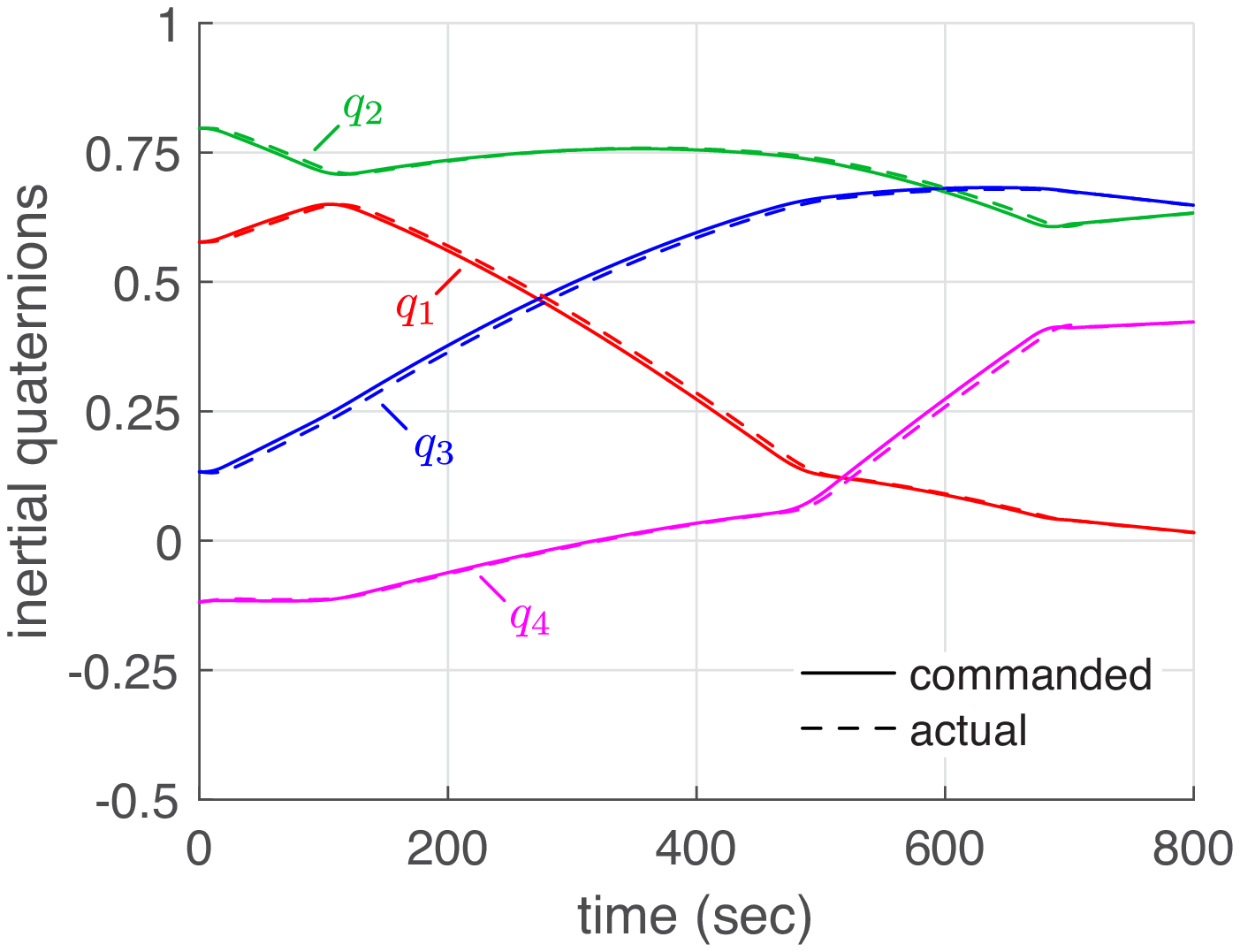}  & 	\includegraphics[width=3.0in,clip]{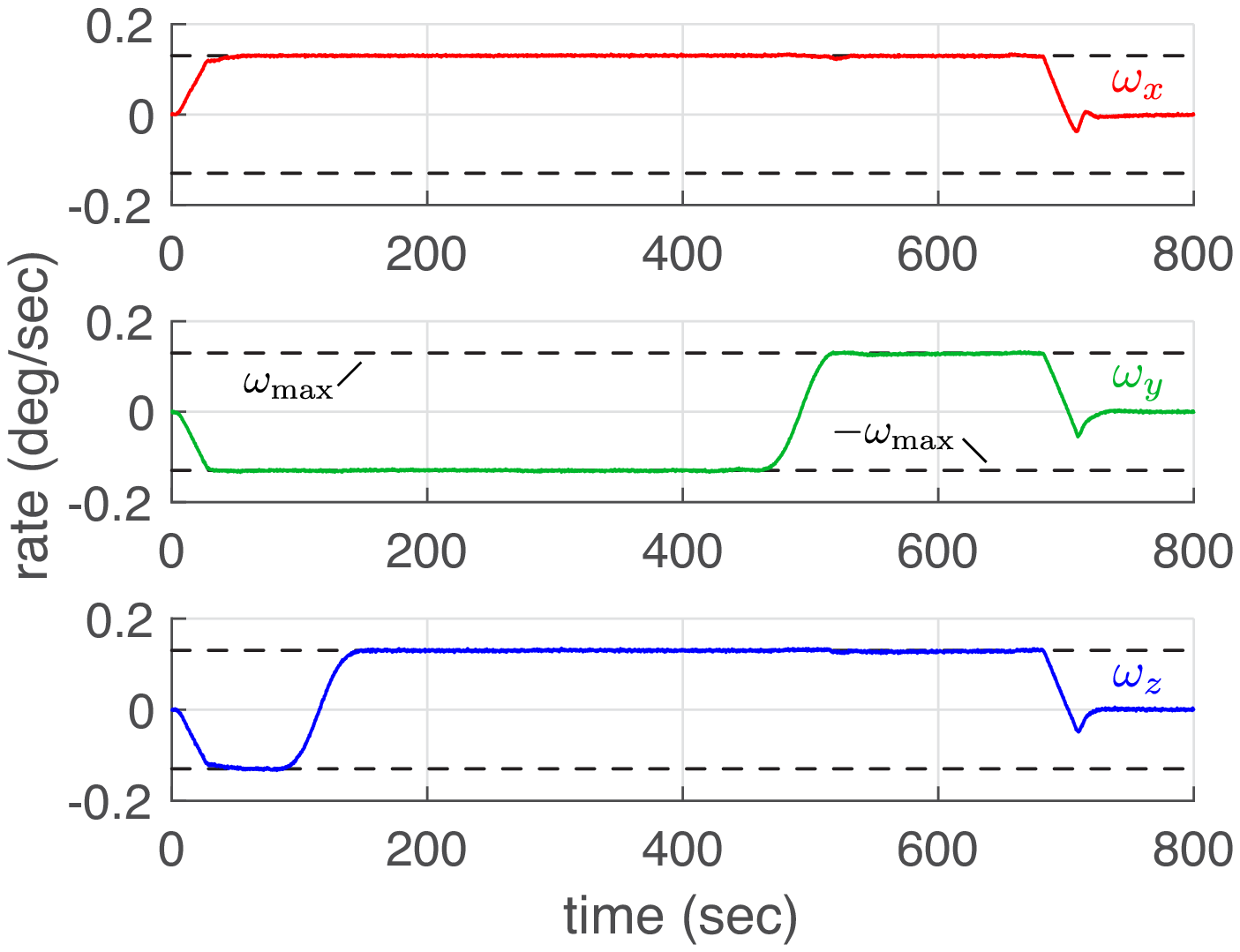} \\
\small{(c)} & \small{(d)} \\
\end{tabular}
\caption{Simulation of a fast attitude maneuver designed for the Lunar Reconnaissance Orbiter: (a) attitude quaternions for waypoint sample-and-hold at the ACS rate; (b) body rates for waypoint sample-and-hold at the ACS rate; (c) attitude quaternions using interpolation filter; (d) body rates using interpolation filter.}
\label{Fig_11}
\end{center}
\end{figure}

The high fidelity simulation results are shown in Figure~\ref{Fig_11}. The response for sample-and-hold at the ACS rate is given in Figures~\ref{Fig_11}a and \ref{Fig_11}b while the response obtained using the Chebyshev filter is given in Figures~\ref{Fig_11}c and \ref{Fig_11}d. The results are nearly indistinguishable for these two cases because each approach can generate input data at the ACS rate. The key difference is that while sample-and-hold requires a data point to be stored for each ACS update, the interpolating filter uses a formula to compute the needed data based on the information encoded in the filter coefficients. The simulation shows that a relatively complex maneuver profile can be reconstructed using only a sparse set of samples.

\newpage
\section{Summary and Conclusion}

Closed-loop attitude steering is an approach for implementing non-standard maneuvers such as fast attitude maneuvers or zero-propellant maneuvers using a conventional spacecraft attitude control system (ACS). Non-standard maneuvers can be executed by loading waypoints into a command buffer for the attitude control system to follow. If uplink bandwidth and data storage are not a concern, the waypoint data may be loaded at the ACS rate for a sample-and-hold implementation. However, command storage capacity is limited for many operational spacecraft and so a downsample-and-hold strategy may be contemplated in order to reduce the transmission and storage requirements. In this paper, we have shown that downsample-and-hold may not be a viable solution as the approach induces unwanted dynamics for quaternion error feedback control. Specifically, downsample-and-hold causes significant ripple in the rate response with peaking that can far exceed the desired slew rate limit. An analysis in the $z$-domain showed that the issue is related to the variation in the DC gain factor of the discrete-continuous attitude control system.
~To resolve the problem of data storage while circumventing the issues associated with downsample-and-hold dynamics, a Chebyshev-like interpolating filter using a Lagrange polynomial basis was proposed. The filter coefficients were used to encode the attitude command trajectories so that an accurate slew trajectory can be reproduced at the ACS rate from the finite impulse response of the filter. Using this approach, closed-loop attitude steering can be implemented with a significantly smaller storage requirements. Simulations using a high fidelity model of the Lunar Reconnaissance Orbiter  as an example of a practical attitude control system illustrated the efficacy of the concept for an example fast maneuver and demonstrated that data storage requirements could be reduced from 70-kilobytes for sample-and-hold at the ACS rate to 1-kilobyte by using the interpolating filter to generate the inputs to the ACS.

\section{Acknowledgement}

The authors would like to thank Prof. I. Michael Ross for his insight and assistance in the formulation and application of the interpolating filter described in this paper. 

\newpage

\section{References}
\vskip 18pt

\begin{spacing}{1.0}

\end{spacing}


\begin{thebibliography}{10}
\newcommand{\enquote}[1]{``#1''}

\bibitem{Wie_book}
Wie, B., \emph{Space Vehicle Dynamics and Control}, AIAA, Reston, VA, 1998.

 \bibitem{Bedrossian_2009}
Bedrossian, N., Bhatt, S., Kang, W., and Ross, I.~M., \enquote{Zero-Propellant Maneuver Guidance,} {\em IEEE Control Systems Magazine\/}, October 2009,  pp.~53--73.
 %
\bibitem{Karpenko_2014}
Karpenko, M., Bhatt, S., Bedrossian, N., and Ross, I. M., ``Flight Implementation of Shortest-Time Maneuvers for Imaging Satellites,'' \emph{Journal of Guidance, Control and Dynamics}, vol. 37, no. 4, pp. 1069-1079, 2014.
%
\bibitem{Lippman_2017}
Lippman, T., Kaufman, J. M., and Karpenko, M., ``Autonomous Planning of Constrained Spacecraft Reorientation Maneuvers,'' \emph{AAS/AIAA Astrodynamics Specialist Conference}, August 20 - 24, 2017 Stevenson, WA. Paper number: AAS 17-676.

\bibitem{Kepler_2015}
Karpenko, M., Ross, I. M., Stoneking, E., Lebsock, K., and Dennehy, N., ``A Micro-Slew Concept for Precision Pointing of the Kepler Spacecraft,'' \emph{AAS/AIAA Astrodynamics Specialist Conference}, August 9 -- August 13, 2015, Vail, CO. Paper number: AAS-15-628.

\bibitem{CMG_patent}
Ross, I. M., Lee, D., and Karpenko, M., \emph{Method and apparatus for contingency guidance of a CMG-actuated spacecraft}, US Patent 9,038,958, issued May 26, 2015.

\bibitem{Marsh_2017}
Marsh, H. C., Karpenko, M., and Gong, Q., ''Electrical-Power Constrained Attitude Steering'', \emph{AAS/AIAA Astrodynamics Specialist Conference}, August 20--24, 2017, Stevenson, WA. Paper number: AAS 17-774.

\bibitem{LRO_breck} 
Karpenko, M., Lippman, T., Ross, I. M., Halverson, J. K., McClanahan, T., Barker, M., Mazarico, E., Dennehy, C. J., Besser, R., VanZwieten, T., and Wolf, A., ``Fast Attitude Maneuvers for the Lunar Reconnaissance Orbiter,'' \emph{42$^{\text{nd}}$ AAS Guidance, Navigation and Control Conference}, February 1--6, 2019, Breckenridge, CO. Paper number: AAS-19-053.

\bibitem{Franklin_1998}
Franklin, G. F., Powell, J. D., and Workman, M. L., \emph{Digital Control of Dynamic Systems}, 3ed., Addison Wesley Longman Inc., Menlo Park, CA, 1998.

\bibitem{Calhoun_2007}
Calhoun, P. C., and Garrick, J. C., ``Observing Mode Attitude Controller for the Lunar Reconnaissance Orbiter,'' \emph{20$^{\text{th}}$ International Symposium on Space Flight Dynamics}, Annapolis, MD, September 24-28, 2007.

\bibitem{IFAC_survey}
Ross, I. M. and Karpenko, M., ``A Review of Pseudospectral Optimal Control: From Theory to Flight,'' \emph{Annual Reviews in Control}, vol. 36, no. 2, pp.182--197, 2012.

\bibitem{Fahroo_2002}
Fahroo, F. and Ross, I. M., ``Direct trajectory optimization by a Chebyshev pseudospectral method,'' \emph{Journal of Guidance, Control, and Dynamics}, vol. 25, no. 1, pp. 160--166, 2002.

\bibitem{Babic_2013}
Babic, D., ``Design of polynomial-based digital interpolation filters based on Chebyshev polynomials,'' \emph{36th International Conference on Telecommunications and Signal Processing}, July 2--4, 2013, Rome, Italy.

\bibitem{DSP_book}
Rawat, T. K., \emph{Digital Signal Processing}, Oxford University Press, New Delhi, India, 2015.

\bibitem{Berrut_2004}
Berrut, J-P. and Trefethen, L. N., ``Barycentric Lagrange Interpolation,'' \emph{SIAM Reveiw}, vol. 46, no. 3, pp. 501--517, 2004.


\end{thebibliography}
\end{document}